\documentclass[aps,prl,twocolumn]{revtex4-1}
\usepackage{amssymb}
\usepackage{epsfig}
\usepackage{graphicx}
\usepackage{mathrsfs}
\usepackage[hang, nooneline, normalsize, FIGTOPCAP]{subfigure}
\usepackage{amsmath}
\usepackage{amsfonts}
\usepackage{mathrsfs}
\newcommand{\ket}[1]{\left\vert #1 \right\rangle}
\newcommand{\bra}[1]{\left\langle #1 \right\vert}

\usepackage{bbold}
\begin{document}

\title{Origin of long-lived oscillations in 2D-spectra of a Quantum Vibronic Model:\\ Electronic vs Vibrational coherence.}

\author{M.~B. Plenio, J. Almeida and S.~F. Huelga }
\address{Institute for Theoretical Physics, Albert-Einstein-Allee 11, University Ulm, D-89069 Ulm, Germany}

\begin{abstract}
We demonstrate that the coupling of excitonic and vibrational motion in biological complexes can provide mechanisms to explain the long-lived oscillations that have been obtained in non linear spectroscopic signals of different photosynthetic pigment protein complexes and we discuss the contributions of excitonic versus purely vibrational components to these oscillatory features. Considering a dimer model coupled to a structured spectral density we exemplify the fundamental aspects of the electron-phonon dynamics, and by analyzing separately the different contributions to the non linear signal, we show that for realistic parameter regimes purely electronic coherence is of the same order as purely vibrational coherence in the electronic ground state. Moreover, we demonstrate how the latter relies upon the excitonic interaction to manifest. These results link recently proposed microscopic, non-equilibrium mechanisms to support long lived coherence at ambient temperatures with actual experimental observations of oscillatory behaviour using 2D photon echo techniques to corroborate the fundamental importance of the interplay of electronic and vibrational degrees of freedom in the dynamics of light harvesting aggregates.
\end{abstract}

\maketitle
\section{Introduction}
Oscillations in the response of a system to external perturbations are a characteristic fingerprint of wave-like, that is, coherent dynamics. However, in the presence of environmental noise, which is inevitable in biological systems, oscillatory features are expected to be short-lived. The observation of persistent oscillations in the non-linear optical spectra of various photosynthetic complexes at both cryogenic and physiological temperatures \cite{greg1,greg2,collini}, together with recent results on individual antenna complexes \cite{vanhulst} and synthetic heterodimers \cite{newgreg}, has led to a multidisciplinary effort to try and identify the specific mechanisms and conditions that enable biological systems to generate such long-lived oscillatory dynamics \cite{felipe,mancal1,ourNP,jonas,mancalSR,cp2013} and to elucidate the functional relevance of such coherent dynamics \cite{MohseniRL+08,PlenioH08,CarusoCD+09,ChinHP12,delReyCH+13,alexandra,scholes2011lessons,thomasreview,cp2013}. \\
These observations are of interest for a variety of reasons. From the point of view of the theory of open quantum systems, the accurate dynamical description of such complexes presents a considerable challenge as it requires transcending the weak coupling and short memory regimes and going beyond merely inducing irreversible decoherence and dissipation. Furthermore, the fact that the observed {\em coherence} persists on time scales that are comparable to the duration of typical transfer time from the absorption in the antennae to the arrival in the reaction center implies the very real possibility that the coherent dynamics of the propagation contributes in a significant way to the (optimal) function of the complex. This in turn would represent a far reaching result which departs from the standard paradigm of relating structure to function by giving a more prominent role to the dynamics \cite{cp2013}.\\
However, any quantitative formulation of such a relation requires as a first step the clear identification of the different coherent mechanisms that contribute to the observed spectral signals. Within this context, a crucial question to answer is: {\em Which dynamical information is being encoded in those oscillatory features?} While initially it was conjectured that long-lived oscillations are a manifestation of purely electronic coherence \cite{greg1,greg2}, this appeared to require unrealistically low reorganization energies of the vibrational environment or the existence of correlated fluctuations, both of which appear unlikely to be correct on the basis of recent numerical work \cite{OlbrichSS+11a,OlbrichSS+11b,ShimRV+12,RengerKS12}. Microscopic models that were more recently proposed to account for the presence of non-transient coherence \cite{felipe,mancal1,ourNP}, however, do require additional vibrational contributions that may end up in the generation of mixed {\em vibronic} coherence or even purely vibrational features \cite{jonas}. \\
The purpose of this work is to analyze the spectral signatures of the model developed in \cite{ourNP}, which explains how picosecond electronic coherence can be driven and supported by quasi-coherent interactions between excitons and spectrally sharp local vibrational modes in the environment (most likely due to intra-molecular motion of the chromophores). The mixing of electronic and vibrational degrees of freedom due to this non-adiabatic dynamics typically results in  vibronic coherence between the dressed states of the electron-phonon coupling. We will show that, for realistic parameter regimes, we expect the resulting overall spectral signal to encompass contributions that have a genuine excitonic component and whose weight is comparable to those resulting from purely vibrational coherence within the electronic ground state manifold. Moreover, we stress again that the manifestation of the latter does require the existence of excitonic coupling, as remarked also in \cite{ourNP} and \cite{jonas}. We emphasize that our results are also compatible with those obtained within the framework of discussing the exciton-phonon interaction in terms of intensity borrowing of dipolar strength \cite{mancalSR} and confirm the relevance of coherent excitonic coupling to explain current spectral observations. \\
We have organized the presentation as follows. Following \cite{ourNP}, a model dimeric structure is presented to capture some of the prominent features displayed by the complete 2D spectral analysis of the Fenna-Matthew-Olson (FMO) complex, which at present is the pigment protein complex which has been subjected to the most detailed investigation. We then discuss the theoretical background underlying the evaluation of the third order response measured in 2D photon echo experiments and clearly identify the different pathways corresponding to processes involving excitonic and vibrational degrees of freedom. Finally we compute exactly the nonlinear response for our model system without resorting to lineshape theory, and evaluate the magnitude of the different paths to assess the relative weight and time scale where excitonic and vibrational coherence manifest.

\section{The Model}
The simplest model system that can give rise to delocalized eigenstates is an exitonically coupled dimer (ecd), consisting of two cofactors, chromophores $a$ and $b$, interacting via an electrostatic Coulomb interaction. The Hilbert space of the two \textit{particles} can be expanded in the site (localized) basis of states $\ket{g}$ (two chromophores de-excited), $\ket{f}$ (both chromophores electronically excited),
$\ket{a}$ (chromophore $a$ excited) and $\ket{b}$ (chromophore $b$ excited). In the case of neutral molecules, the Coulomb potential between different chromophores is dominated by their dipole-dipole interaction and the system Hamiltonian can be expressed as
\begin{equation}
    H_{\rm ecd}\equiv \frac{\mathcal{E}_a}{2} \sigma_a^z + \frac{\mathcal{E}_b}{2} \sigma_b^z
    + J(\sigma_a^+ \sigma_b^- + \sigma_b^- \sigma_a^+)
    \label{eq:Hecd}
\end{equation}
where $\mathcal{E}_a$ and $\mathcal{E}_b$ are the energy gap of cofactors $a$ and $b$ respectively
and $J = [\vec{\mu}_a \cdot \vec{\mu}_b - 3(\vec{\mu}_a \cdot \vec{R}/R)(\vec{\mu}_b \cdot \vec{R}/R)]/R^3$ is the standard dipole-dipole interaction energy between two point electric dipoles $\vec{\mu}_a$, $\vec{\mu}_b$ residing on cofactors connected by a vector $\vec{R}$ of length $R$.\\

The Hamiltonian eq. (\ref{eq:Hecd}) conserves the number of electronic excitations. This implies that both the ground state $\ket{g}$ and the doubly excited states $\ket{f}$ are stationary states with eigenenergies $E^{\rm ecd}_{g,f}=\mp(\mathcal{E}_a+\mathcal{E}_b)/2.$ On the other hand, the excitonic coupling $J$ allows for transitions between singly excited states $\ket{a}$ and $\ket{b}$ giving rise to eigenstates states $\ket{A}$ and $\ket{B}$ which will generally correspond to a delocalized excitation. The exact form of these eigenvectors can be expressed in terms of the mixing angle $\theta$, defined implicitly as $ {\rm tan}(2\theta)=2J/\mathcal{E}_{ab},$ where $\mathcal{E}_{ab}\equiv \mathcal{E}_b-\mathcal{E}_a$. With this definition, the eigenstates $\ket{A}$ and $\ket{B}$ in the one-excitation sector of the Hamiltonian Eq.(\ref{eq:Hecd}) can be written as
\begin{equation}
  \left(
  \begin{array}{c}
  \ket{A}\\ \ket{B}
  \end{array}
  \right)
  =
  \left(
  \begin{array}{cc}
  \cos(\theta)& \sin(\theta) \\
  -\sin(\theta)& \cos(\theta)
  \end{array}
  \right)
  \left(
  \begin{array}{c}
  \ket{a}\\ \ket{b}
  \end{array}
  \right).
  \label{eq:ecd_eigenvectors}
\end{equation}
These expressions make explicit the degree of delocalization between both chromophores of the new
eigenstates (excitons) as a function of the mixing angle $\theta$. The exciton energies are split by $\Delta_{\rm ex}=E^{\rm ecd}_B - E^{\rm ecd}_A,$ where $E^{\rm ecd}_{A,B} = \mp (1/2)\sqrt{\mathcal{E}_{ab}^2+4J^2}.$\\
The vibrational environment of real protein-pigment complexes is characterized by a highly structured spectral function that encompasses both smooth and sharp features \cite{marcus}. The coupling of excitonic degrees of freedom to vibrational modes with energies in the vicinity of the excitonic splitting $\Delta_{\rm ex}$ can have a significant impact on the electronic dynamics, as discussed in \cite{ourNP}.  In particular, this coupling can express itself in the persistence of oscillatory features in observables pertaining to purely excitonic degrees of freedom such as the survival probability of a selected excitonic superposition state \cite{ourNP}. Here we are interested in studying explicitly signatures of these vibrational modes in non-linear photon echo signals and, more specifically, in analyzing the effect that these modes have on the dynamics of the peak beatings in the population time. To this effect we will extend our Hamiltonian Eq.(\ref{eq:Hecd}) to include the degrees of freedom of two identical normal modes coupled to each chromophore as
\begin{equation}
    H_{\rm e-v} = H_{\rm ecd} + \!\!\sum_{k=a,b} \!\omega a_k^{\dagger}a_k^{\, }
    + \tfrac{\omega}{2} \sqrt{S_{\rm HR}} \sum_{k=a,b}\!\sigma^z_k (a_k^{\dagger} + a_k),
    \label{eq:Htot}
\end{equation}
where $\omega$ stands for the frequency of the normal mode, $a_k^{\dagger}$, $a_k$, ($k=a,b$) denote the phonon creation and annihilation operators on the corresponding vibrational modes and $S_{\rm HR}$ denotes the Huang-Rhys factor \cite{ratsep2007electron} that scales the linear electron-phonon coupling. \\
\begin{figure}[hbt]
\includegraphics[width=8cm]{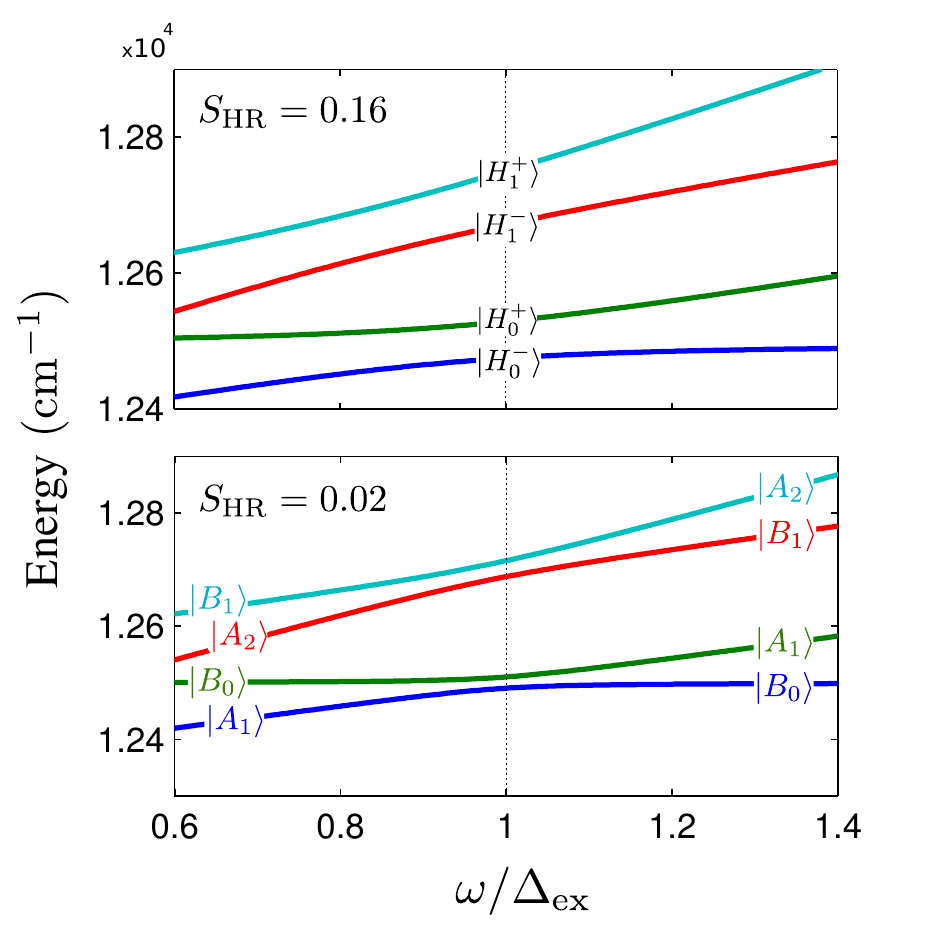}
\caption{\textbf{Self avoided energy level crossings and state hybridization in the one-exciton manifold.}  Energies have been obtained by diagonalising the vibronic dimer described by eq.~\eqref{eq:Htot} for two different values of the Huang-Rhys factor $S_{\rm HR}=0.02$ and $S_{\rm HR}=0.16$ and the parameters described in the text. The interaction with a vibrational mode with energy close to the electronic splitting $\omega\simeq\Delta_{\rm ex}$ produces a hybridization of the states $\ket{A_1}$ and $\ket{B_0}$ resulting in a new pair of states $\ket{H_0^{\pm}}=\cos(\phi)\ket{A_1}\pm\sin(\phi)\ket{B_0}$ with the angle $\phi$ increasing from $0$ to $\pi/2$ as we cross the resonance point from lower to higher values of $\omega/\Delta_{\rm ex}$. Mixing between $\ket{A_1}$ and $\ket{B_0}$ is therefore quite significant close to the resonant point $\omega/\Delta_{\rm ex}=1$. An analogous reasoning can be extended to the hybrid states $\ket{H_1^{\pm}}$ resulting from the coupling between states $\ket{A_2}$ and $\ket{B_1}$. State hybridization allows the sharp dipolar transitions between states with different vibrational numbers depicted in fig.~\ref{fig:dipolar_transitions}c. The energy splitting between these hybrid states is seen from these graphs to grow with increasing interaction strength $S_{\rm HR}$.}
\label{fig:level_crossing}
\end{figure}
In order to keep the number of parameters of the model as small as possible while still illustrating the fundamental physics, we will consider initially a dimer system where only the higher energy chromophore is subject to a vibrational coupling. This assumption will be relaxed later when 2D spectra are evaluated. In the following we will consider a model system whose parameters are inspired by the dominant dimeric structure in FMO, as provided by sites 3 and 4 \cite{fmo}. The site energies for each chromophore are $\mathcal{E}_{a}=12328$ ${\rm cm^{-1}}$ and $\mathcal{E}_{b}=12472$ ${\rm cm^{-1}}$ and the dipolar coupling is chosen as $J=70.7$ ${\rm cm^{-1}}$. The resulting excitonic splitting is $\Delta_{\rm ex}=202$ ${\rm cm^{-1}}$. With these numbers the difference in energy between excitons $\ket{A}$ and $\ket{B}$ with respect to the electronic ground state $\ket{g}$ are $E^{\rm ecd}_A-E^{\rm ecd}_g=12299$ ${\rm cm^{-1}}$ and $E^{\rm ecd}_B-E^{\rm ecd}_g=12501$ ${\rm cm^{-1}}$. It is convenient to express the parameters of our system in terms of the physically relevant exciton energy splitting $\Delta_{\rm ex}$
of our purely electronic Hamiltonian~\eqref{eq:Hecd}. This yields $\mathcal{E}_{ab}=0.70 \Delta_{\rm ex}$, $J=-0.35\Delta_{\rm ex}$ and a mixing angle $\theta=\pi/8$, which corresponds to a significant degree of delocalization of the electronic eigenstates over both cofactors. The spectrum of the total Hamiltonian eq.~\eqref{eq:Htot} can not be fully specified in terms of the eigenstates of the purely electronic
dimer $H_{\rm ecd}$. The exciton-vibrational coupling renders the eigenstates of the full Hamiltonian a complex superposition of electronic and vibrational wavefunctions. Nonetheless, if we constrain ourselves to a regime where the interaction strength with the harmonic mode is moderate, the structure of eigenstates of the full Hamiltonian can still be understood as different vibrational progressions of each electronic state of the purely electronic dimer. In this scenario we will denote the states $\ket{g_n}$, $\ket{A_n}$ and $\ket{B_n}$ the corresponding vibrational progression of states corresponding to the ground-state electronic manifold and the first and second excitonic states. Here the index $n$ increases with increasing energy states within the vibrational progression.
In the next sections we will compute the non-linear optical response
of the model described by Hamiltonian eq.\eqref{eq:Htot}. The effect of
the coupling between vibrational and electronic degrees of freedom
will immediately lead to two phenomena that will leave
long-lived beating traces in the population time. One of these
vibrationally-induced effects is expressed in the one-exciton sector
of our model while the other concerns entirely the ground state
electronic manifold. In order to understand the latter we
have to carefully examine the spectrum of the system in the
one-exciton manifold. In fig.~\ref{fig:level_crossing} we have
obtained some relevant energies diagonalising the vibronic dimer
described by Eq.~\eqref{eq:Htot}. The interaction with a vibrational
mode with energy close to the electronic splitting
$\omega\simeq\Delta_{\rm ex}$ produces a hybridization of the states
$\ket{A_1}$ and $\ket{B_0}$ resulting in a new pair of states
\begin{equation}
    \ket{H_0^{\pm}}=\cos(\phi)\ket{A_1}\pm\sin(\phi)\ket{B_0}
\end{equation}
with the angle $\phi$ increasing from $0$ to $\pi/2$ as we cross the resonance
point from lower to higher values of $\omega/\Delta_{\rm ex}$. In
particular, mixing between states $\ket{A_1}$ and $\ket{B_0}$ is quite
significant in the vicinity of the resonant point $\omega/\Delta_{\rm ex}=1$. An
analogous reasoning can be extended to the hybrid states
$\ket{H_1^{\pm}}$ resulting from the coupling between states $\ket{A_2}$ and
$\ket{B_1}$. This mixing is of crucial importance to understand how different
(electronic vs vibrational) coherent contributions are mapped out in
2D spectra. Indeed, level hybridization will be at the heart of the sharp
dipolar transitions between states with different vibrational numbers depicted
in fig.~\ref{fig:dipolar_transitions}c which, as well will explain in detail,
result in the pathways depicted in fig.~\ref{fig:Feynman_diagrams}b.

\begin{figure}
\includegraphics[width=\columnwidth]{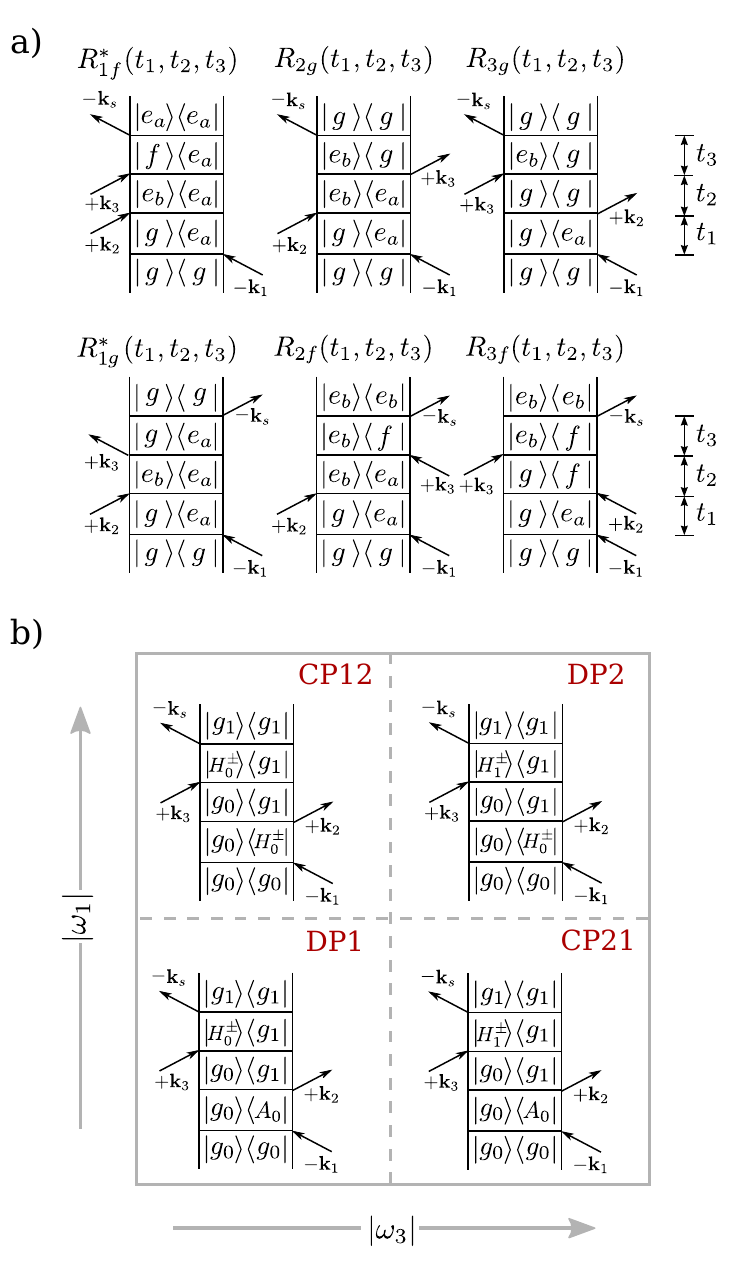}
\caption{\textbf{Contributions to the 2D electronic photon echo signal.}  Diagrammatic representation of the processes involved in the total signal measured in a 2D electronic photon echo experiment in the spatial direction $\textbf{k}_s=-\textbf{k}_1+\textbf{k}_2+\textbf{k}_3$. In the impulsive limit, within the rotating wave approximation and assuming strict ordering of the laser pulses such that $\textbf{k}_1$, $\textbf{k}_2$ and $\textbf{k}_3$ are the propagation directions of the first, second and third pulses that interact with the sample, the only diagrams that contribute to the signal are $R^{*}_{1f}(t_1, t_2, t_3)$, $R_{2g}(t_1, t_2, t_3)$ and $R_{3g}(t_1, t_2, t_3)$. For the sake of completeness in panel \textit{a)} we have plotted all the processes contained in the functions $R_{1}(t_1, t_2, t_3)$, $R_{2}(t_1, t_2, t_3)$ and $R_{3}(t_1, t_2, t_3)$ of a purely electronic dimer. The first row contains the rephasing contributions involved in the total signal while the second row contains non-rephasing contributions that are negligible under the particular conditions expressed above. In panel \textit{b)} we plot another set of processes included in the $R_{3g}(t_1, t_2, t_3)$ contribution of an electronic dimer interacting with a vibrational mode as explained in the text. These four sub-diagrams result in vibrational superpositions of states within the ground state manifold during the population time $t_2$ of the experiment. The position of the diagrams in the figure is in correspondence with the cross-diagonal
(CP12, CP21) and diagonal (DP1, DP2) peak amplitude to which they contribute.}
\label{fig:Feynman_diagrams}
\end{figure}

\section{Dynamical equations}
Our core interest is the study of the population time beatings in the various peaks of the 2D spectrum of the vibronic dimer defined in eq. \eqref{eq:Htot} as well as the mechanisms underlying their unexpectedly long lifetimes. With this in mind, rather than considering a detailed microscopic description of the dephasing processes involved in a real photosynthetic complex we will consider a simplified model of the environment which, nevertheless, contains those features of the full description that are necessary to understand the origin of long-lasting population time beatings.

In order to describe the dynamical evolution of our system we will therefore use a Markovian master equation where the effect of the electronic dephasing will be explicitly included with the appropriate rate $\gamma_{\rm deph}$ {\em and} we we include vibrational modes that couple both to the electronic degrees of freedom and are additionally coupled dissipatively to a Markovian environment at a finite temperature. This results in equations of motion given by
\begin{eqnarray}
    \frac{d\rho}{dt}&=&\mathscr{L}(\rho), \text{ with }\\
    \mathscr{L}(\rho)&\equiv& -i/\hbar [H_{\rm e-v}, \rho]
    +\gamma_{\rm deph}( \sigma^z_a \rho \sigma^z_a + \sigma^z_b \rho \sigma^z_b - 2\rho )\nonumber\\
    && + \gamma_{mod}(n_T+1)\sum_{k=a,b} [-a^{\dagger}_k a_k \rho - \rho a^{\dagger}_k a_k + 2a_k\rho a^{\dagger}_k]\nonumber\\
    && + \gamma_{mod}n_T\sum_{k=a,b} [-a_k a^{\dagger}_k \rho - \rho a_k a^{\dagger}_k + 2a^{\dagger}_k\rho a_k]
    \label{eq:ME}
\end{eqnarray}
where $n_T$ is the mean thermal occupation number of the vibrational mode and $\gamma_{mod}$ is the damping rate into the Markovian thermal reservoir to which the vibrational modes are coupled.

The action of the operator $\mathscr{L}(\rho)$ is manifestly linear in the density matrix and it is convenient to move to a tetradic representation where the action of superoperators can be written as ordinary matrix multiplications. In tetradic notation the master equation above can be rewritten as
\begin{equation}
\frac{d\vert{\rho}\rangle\rangle}{dt}=\mathscr{L}\vert{\rho}\rangle\rangle,
\label{eq:ME_tetradic}
\end{equation}
where the tetradic representation $\vert A \rangle\rangle$ of an
ordinary matrix operator $A$ of dimensions $n\times n$ is a vector of
dimensions $n^2\times 1$ consisting on the $n$ columns of the matrix
$A$ written one below the other. Note also that we have used the
same letter $\mathscr{L}$ to denote both the Liouvillian operator in
eq. \eqref{eq:ME} and its matrix superoperator representation in
eq. \eqref{eq:ME_tetradic}. This ambiguity is however removed with the
use of the tetradic notation with double brackets as in the equation
above. With this prescription it is straightforward to obtain the matrix
form of the superoperator $\mathscr{L}$ in the equation above by making
use of the following algebraic identity
\begin{equation}
    \vert A\rho B\rangle\rangle = (B^{t}\otimes A)\vert \rho \rangle\rangle.
\end{equation}
Here $A$, $\rho$ and $B$ are ordinary matrices. The advantage of this notation
is clear from eq.~\eqref{eq:ME_tetradic} since we can now formally solve the
equation of motion to obtain
\begin{equation}
    \vert \rho(t) \rangle\rangle = \mathscr{G}(t)\vert \rho(0) \rangle\rangle,
    \label{eq:defGreenfunction}
\end{equation}
with the propagator
\begin{equation}
    \mathscr{G}(t)\equiv e^{\mathscr{L}t}.
\end{equation}
%

\section{Electronic 2D photon echo spectroscopy}
In this section we describe how to compute the 2D photon echo spectroscopy signal in the joint perturbative and impulsive limit which applies to a wide variety of experiments such as \cite{greg1,greg2,collini,newgreg}. The basic setup involves the illumination of the system by three consecutive, ultrashort laser pulses \cite{echo,echo1,echo2}. The interaction between each laser pulse and the vibronic dimer (eq.~\eqref{eq:Htot}) is described by a dipolar coupling of the form $H_{\rm int}(t)=V\cdot E(\textbf{r}, t)$, where the operator $V$ denotes the total electric-dipole operator of the system, $V=\vec{\mu}_a\sigma^x_a+\vec{\mu}_b\sigma^x_b$. The total electric field $E(\textbf{r}, t)$ can be parametrized as
\begin{align}
E(\textbf{r}, t)&=\sum_{r=1}^3 \vec{\mathcal{E}}_r G(t-t^0_r, \Delta_r)E_0 \sin(\omega_r(t-t_r)+\textbf{k}_r\textbf{r}),
\label{eq:defHint}
\end{align}
where $\vec{\mathcal{E}}_r$ is the polarization vector of each pulse and $E_0$ is the field strength. The function $G_r(t, \Delta)$ stands for some pulse envelope function and $\Delta$ is a measure of its width. The time variables $t^0_r$ indicate the moment in time where each pulse is acting. In 2D photon echo experiments the laser polarization of the successive pulses is usually taken to be independent of $r$, $\vec{\mathcal{E}}=\vec{\mathcal{E}}_r$ and we can absorb its action in the definition of the total electric-dipole operator
\begin{equation}
    V=\mu_a\sigma^x_a+\mu_b\sigma^x_b, \;\;\text{ with }
    \mu_a\equiv \vec{\mu}_a\cdot\vec{\mathcal{E}},\;
    \mu_b\equiv \vec{\mu}_b\cdot\vec{\mathcal{E}}.
\end{equation}
The electric field measured in a non-linear electronic 2D photon echo experiment is essentially proportional to the polarization
$P(t)=\langle V\rho(t)\rangle$ induced in the sample by the incoming
pulses. When this polarizarion is measured in the particular spatial
direction $\textbf{k}_s=-\textbf{k}_1+\textbf{k}_2+\textbf{k}_3$ and
under low laser intensity, the dynamics can be evaluated perturbatively and the dominant term comes from the third order contribution, which can be formally expressed as
\begin{multline}
P^{(3)}(t)\equiv \langle V\rho^{(3)}(t)\rangle
= \int_{0}^{\infty}dt_3\int_{0}^{\infty}dt_2\int_{0}^{\infty}dt_1 \\
S^{(3)}(t_1, t_2, t_3)
E(\textbf{r}, t-t_3)E(\textbf{r}, t-t_3-t_2)E(\textbf{r}, t-t_3-t_2-t_1),
\label{eq:defP3}
\end{multline}
where the third-order response function $S^{(3)}(t_1,t_2,t_3)$ can be
written as the sum of four terms
\begin{multline}
S^{(3)}(t_1,t_2,t_3)=
R_1(t_1,t_2,t_3)+R_2(t_1,t_2,t_3)+\\
+R_3(t_1,t_2,t_3)+R_4(t_1,t_2,t_3) -{\rm c.c.}.
\end{multline}
Using tetradic notation the form of the functions above can be reexpressed
in a very compact form:
\begin{align}
R_1(t_1, t_2, t_3)&=
\langle\langle V\vert \mathscr{G}(t_3)\overrightarrow{\mathscr{V}}
\mathscr{G}(t_2)\overrightarrow{\mathscr{V}}
\mathscr{G}(t_1)\overleftarrow{\mathscr{V}}
\vert \rho(-\infty)\rangle \rangle \nonumber\\
R_2(t_1, t_2, t_3)&=
\langle\langle V\vert \mathscr{G}(t_3)\overrightarrow{\mathscr{V}}
\mathscr{G}(t_2)\overleftarrow{\mathscr{V}}
\mathscr{G}(t_1)\overrightarrow{\mathscr{V}}
\vert \rho(-\infty)\rangle \rangle \nonumber\\
R_3(t_1, t_2, t_3)&=
\langle\langle V\vert \mathscr{G}(t_3)\overleftarrow{\mathscr{V}}
\mathscr{G}(t_2)\overrightarrow{\mathscr{V}}
\mathscr{G}(t_1)\overrightarrow{\mathscr{V}}
\vert \rho(-\infty)\rangle \rangle \nonumber \\
R_4(t_1, t_2, t_3)&=
\langle\langle V\vert \mathscr{G}(t_3)\overleftarrow{\mathscr{V}}
\mathscr{G}(t_2)\overleftarrow{\mathscr{V}}
\mathscr{G}(t_1)\overleftarrow{\mathscr{V}}
\vert \rho(-\infty)\rangle \rangle, \nonumber \\
\label{eq:defS3}
\end{align}
with the action of the superoperators $\overleftarrow{\mathscr{V}}$
and $\overrightarrow{\mathscr{V}}$ upon an ordinary operator
$A$ defined as $\overleftarrow{\mathscr{V}}A\equiv VA$,
$\overrightarrow{\mathscr{V}}A\equiv AV$ and with $\langle\langle
A\vert B\rangle\rangle\equiv {\rm Tr}\{A^{\dagger}B\}$.

The form of the third order polarization adopts a simpler form under
the following approximations: i/ Rotating wave approximation in
eq.~\eqref{eq:defP3}, ii/ impulsive limit, i.e., $\Delta_k \to 0$ in
the electric field definition in eq.~\eqref{eq:defHint} and iii/
strict time ordering of the pulses, such that $t^0_1<t^0_2<t^0_3$ also
in eq.~\eqref{eq:defP3}. With these assumptions, the form of the 2D
photon echo signal measured in the spatial direction
$\textbf{k}_s=-\textbf{k}_1+\textbf{k}_2+\textbf{k}_3$ involves only three contributions:
\begin{multline}
P^{(3)}(t_1, t_2, t_3)\simeq E_0^3
\Big( -R_{1f}^{*}(t_1, t_2, t_3) +\\
+ R_{2g}(t_1, t_2, t_3) +R_{3g}(t_1, t_2, t_3)\Big),
\end{multline}
where we have renamed the time variables such that $t_1\equiv t^0_2-t^0_1$
(time separation between the second and first pulses),
$t_2\equiv t^0_3-t^0_2$ (time separation between third and second pulses)
and $t_3\equiv t-t^0_3$ (time separation between the actual measurement and
the third pulse). The functions $R_{1f}(t_3, t_2, t_1)$, $R_{2g}(t_3,
t_2, t_1)$ and $R_{3g}(t_3, t_2, t_1)$ correspond to the subset of
processes represented in fig.~\ref{fig:Feynman_diagrams}a. Indeed
these pathways can be properly selected modifying
eqs.~\eqref{eq:defS3} such that the dipole operator acting after each
time contain only the matrix elements taking place in the particular
pathway. That is,
\begin{align}
R_{1f}(t_1, t_2, t_3)&=
\langle\langle V\vert \mathscr{G}(t_3)\overrightarrow{\mathscr{V}}_{ef}
\mathscr{G}(t_2)\overrightarrow{\mathscr{V}}_{ge}
\mathscr{G}(t_1)\overleftarrow{\mathscr{V}}_{ge}
\vert \rho(-\infty)\rangle \rangle \nonumber\\
R_{2g}(t_1, t_2, t_3)&=
\langle\langle V\vert \mathscr{G}(t_3)\overrightarrow{\mathscr{V}}_{ge}
\mathscr{G}(t_2)\overleftarrow{\mathscr{V}}_{ge}
\mathscr{G}(t_1)\overrightarrow{\mathscr{V}}_{ge}
\vert \rho(-\infty)\rangle \rangle \nonumber\\
R_{3g}(t_1, t_2, t_3)&=
\langle\langle V\vert \mathscr{G}(t_3)\overleftarrow{\mathscr{V}}_{ge}
\mathscr{G}(t_2)\overrightarrow{\mathscr{V}}_{ge}
\mathscr{G}(t_1)\overrightarrow{\mathscr{V}}_{ge}
\vert \rho(-\infty)\rangle \rangle, \nonumber \\
\label{eq:defS3_reduced}
\end{align}
with the new superoperators $\overrightarrow{\mathscr{V}}_{ge}$, $\overrightarrow{\mathscr{V}}_{ef}$ (and
$\overleftarrow{\mathscr{V}}_{ge}$, $\overleftarrow{\mathscr{V}}_{ef}$) defined from the truncated dipolar operators $V_{ge}$ and $V_{ef}$, whose form is the same as the complete operator $V$ but retaining only the matrix elements connecting the ground-state and one-exciton-manifolds (for $V_{ge}$) and the elements conecting the one-exciton- and doubly-excited-manifolds (for $V_{ef}$).
For the following we would like to note that it will be advantageous to evaluate the expressions eq.\ref{eq:defS3_reduced} not in $t_1$ and $t_3$ but directly in Fourier space by carrying out the Fourier transform on the propagators $\mathscr{G}(t_1)$ and $\mathscr{G}(t_3)$ analytically to obtain

\begin{equation}
\mathscr{\tilde G}(\omega_1) := \int_0^{\infty} e^{-i\omega_1 t} \mathscr{G}(t) = \frac{1}{i\omega_1 -
\mathscr{L}}
\end{equation}

to find
\begin{align}
R_{1f}(\omega_1, t_2, \omega_3)&=
\langle\langle V\vert \mathscr{\tilde G}(\omega_3)\overrightarrow{\mathscr{V}}_{ef}
\mathscr{G}(t_2)\overrightarrow{\mathscr{V}}_{ge}
\mathscr{\tilde G}(\omega_1)\overleftarrow{\mathscr{V}}_{ge}
\vert \rho(-\infty)\rangle \rangle \nonumber\\
R_{2g}(\omega_1, t_2, \omega_3)&=
\langle\langle V\vert \mathscr{\tilde G}(\omega_3)\overrightarrow{\mathscr{V}}_{ge}
\mathscr{G}(t_2)\overleftarrow{\mathscr{V}}_{ge}
\mathscr{\tilde G}(\omega_1)\overrightarrow{\mathscr{V}}_{ge}
\vert \rho(-\infty)\rangle \rangle \nonumber\\
R_{3g}(\omega_1, t_2, \omega_3)&=
\langle\langle V\vert \mathscr{\tilde G}(\omega_3)\overleftarrow{\mathscr{V}}_{ge}
\mathscr{G}(t_2)\overrightarrow{\mathscr{V}}_{ge}
\mathscr{\tilde G}(\omega_1)\overrightarrow{\mathscr{V}}_{ge}
\vert \rho(-\infty)\rangle \rangle. \nonumber \\
\label{eq:defS3_reduced}
\end{align}

\begin{figure*}
\includegraphics[height=5cm]{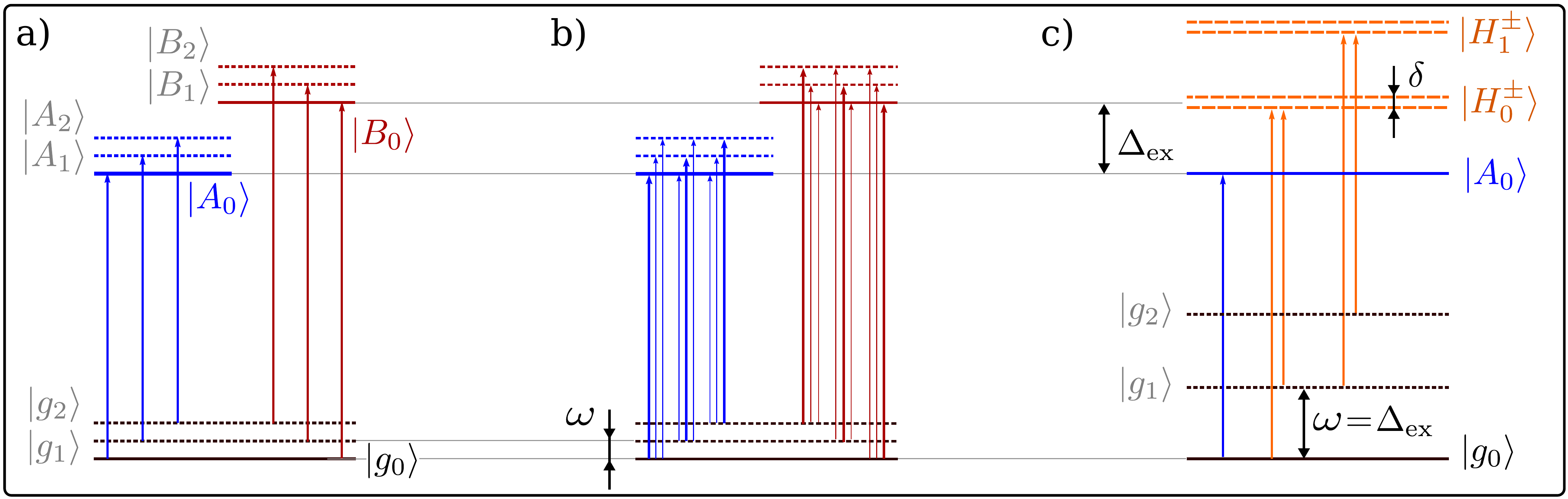}
\caption{\textbf{Dipolar transition strengths in the electronic-vibrational dimer.} Schematic representation of the electric-dipolar transitions in the model described by eq.\eqref{eq:Htot}. \textit{a) low electronic-vibrational coupling regime.} In the limit of low Huang-Rhys factor the total
eigenstates of the system can be well approximated by the corresponding tensor product of the eigenstates of the electronic dimer and the harmonic oscillator. Under this assumption the electric-dipole operator can only connect states on different electronic manifolds and with the same vibrational quantum number. This situation is depicted with the single intense transition lines represented in the scheme. \textit{b) intermediate electronic-vibrational coupling regime}. In the regime of intermediate Huang-Rhys factor the potential energy surfaces of each electronic manifold are displaced with respect to each other. In this situation the electric-dipole operator can have non-zero matrix elements connecting eigenstates with different vibrational numbers. The value of these matrix elements is proportional to the so called Franck-Condon overlap between the vibrational part of the involved wavefunctions. These dipolar transitions weighted by the corresponding Franck-Condon overlap are represented pictorially in the figure with lines of different thickness. In the scheme of the figure we have assumed that the interaction with the mode is low enough to preserve the delocalized excitonic nature of the eigenfunctions. \textit{c) resonant electronic-vibrational regime.} The presence of a harmonic mode resonant with the excitonic splitting results in a particular hybridization of states with different vibrational numbers (see text). The resulting hybrid states in the one-exciton manifold $\ket{H_0^{\pm}}$ and $\ket{H_1^{\pm}}$ can act, even in the limit of weak electronic-vibrational coupling, as a bridge between states in the ground state manifold with different vibrational numbers, therefore behaving as a {\em catalyzer} for sustaining (purely vibrational) coherent oscillations within the electronic ground state manifold.}
\label{fig:dipolar_transitions}
\end{figure*}

\section{Results and Discussion}
From eqs.~\eqref{eq:defS3_reduced} it can be seen that the output of a spectroscopic signal is dictated by, on the one hand, the particular dynamical evolution of the system and, on the other hand, the dipolar momenta of the system's optical transitions. The dynamical evolution of the system is contained in the propagator $\mathscr{G}(t)$, while the strength of the dipolar transitions is given by the matrix elements of the electric-dipole operator $V$. In order to observe long-lived dynamical features of any kind during the population time of an electronic 2D experiment we require: i/ an intrinsic dynamical behavior during this period of time consistent with these long dephasing times and ii/ significant amplitude of the different dipolar transitions that play a role in this particular process. Resorting to a simplified diagrammatic viewpoint appropriate to the perturbative setting that we are considering here, the amplitude of a certain spectroscopic complete pathway is proportional to the product of the four different matrix elements of the dipolar operator involved in the diagram (see fig 2) describing this particular process. The interaction of a purely electronic dimer with localized vibrational modes, as described by the Hamiltonian $H_{\rm e-v}$ in eq.~\eqref{eq:Htot}, leads to crucial changes in the two aforementioned factors and therefore the system's response can differ significantly from that of a purely excitonic coupled dimer \cite{tonu}. \\
In particular, long-lasting beatings of the electronic 2D spectra signal during the population time may be facilitated by the vibronic coupling \cite{ourNP}. This mechanism can be cast in terms of two interacting systems with intrinsically different dephasing time scales: a purely vibrational superposition tends to exhibit long lifetimes while the life time of a superposition of exciton states is subject to rapid dephasing due to the relatively strong interaction with the broad band environment. In a (relatively) weak electronic-vibrational coupling regime, the mode experiences an effective dephasing rate mediated by the electronic system. This effective decoherence time can in general be much longer that the excitonic dephasing times. Any electronic-vibrational superposition of states during the population time can hence show extended dephasing times. Note, however, that it is still not immediately transparent how this effect will manifest when the system is probed via a photon echo sequence as a quasi-resonant coupling between electronic and vibrational motion strongly affects the structure of the energy levels in the one-exciton sector, as shown in fig.~\ref{fig:level_crossing}. The existence of self-avoided level crossings in the energy level structure is of crucial importance in the redistribution of the strength between dipolar transitions involving the ground-state- and the one-exciton-manifolds and will therefore dictate the characteristics of the optical response.\\
In order to understand the main effects, and for the sake of clarity, we will restrict the discussion to a weak electronic-vibrational interaction regime. Let us begin with the scenario in which vibrational mode and excitonic energy differences are non-resonant, in which case we describe the eigenstates of our system, to first order, by separable states containing an electronic wavefunction and the corresponding vibrational one. That is, product states of the form $\ket{G_N}\simeq \ket{G}\ket{\xi_{G,N}}$ (and analogously for the states $\ket{A_N}$ and ${B_N}$). In this situation the transition strength between the excitonic states $\ket{A_N}$, $\ket{B_N}$ and the ground state $\ket{G_M}$ is weighted by the vibrational Franck-Condon factor $\bra{\xi_ {G,M}}\xi_{a,N}\rangle$. Roughly speaking, these factors quantify the overlap of two displaced harmonic oscillator wavefunctions with respect to a fixed reference. If we take the displacement (related to the parameter $d$ in our model) tend to zero, the overlap $\bra{\xi_ {G,M}}\xi_{a,N}\rangle$ can be seen to behave as a delta function giving a finite value only when $M=N$. That is, the dipolar transition strength is only different from zero between states with the same vibrational quantum number (see fig.~\ref{fig:dipolar_transitions}a). Finite values of the displacement $d$ result in a small but finite overlap and hence transition strength between states with different vibrational quantum numbers. We emphasize that these dipolar strengths between different vibrational states tend to be weaker than those transitions connecting states with the same vibrational quantum numbers (see fig.~\ref{fig:dipolar_transitions}b where we have used thinner lines to represent the weaker transitions between different vibrational states). \\
An important regime concerning the distribution of dipolar transition strength different from the cases discussed above is found when a resonance between excitonic transition frequencies and vibrational frequencies occurs. In this situation a description via electronic-vibrational product states is no longer appropriate and it is advantageous to adopt a description in terms of the dressed states $\ket{H_0^{\pm}}$. The dipolar transition strength between the states $\ket{H_0^{\pm}}$ and the states $\ket{g_0}$ and $\ket{g_1}$ in the ground state manifold may indeed have comparable values even though the transitions involve different quantum vibrational numbers. This is easy to see from the definition of the dipolar transition strength
\begin{multline}
\bra{H_0^{\pm}}\hat{\mu}\ket{g_0}=\cos(\phi)\bra{A_1}\hat{\mu}\ket{g_0}\pm\\
\pm \sin(\phi)\bra{B_0}\hat{\mu}\ket{g_0}
\simeq
\sin(\phi) \mu_B
\end{multline}
and
\begin{multline}
\bra{H_0^{\pm}}\hat{\mu}\ket{g_1}=\cos(\phi)\bra{A_1}\hat{\mu}\ket{g_1}\pm\\
\pm \sin(\phi)\bra{B_0}\hat{\mu}\ket{g_1}
\simeq
\cos(\phi) \mu_A,
\end{multline}
\begin{figure}
\vspace*{-3.cm}
\includegraphics[scale=0.37]{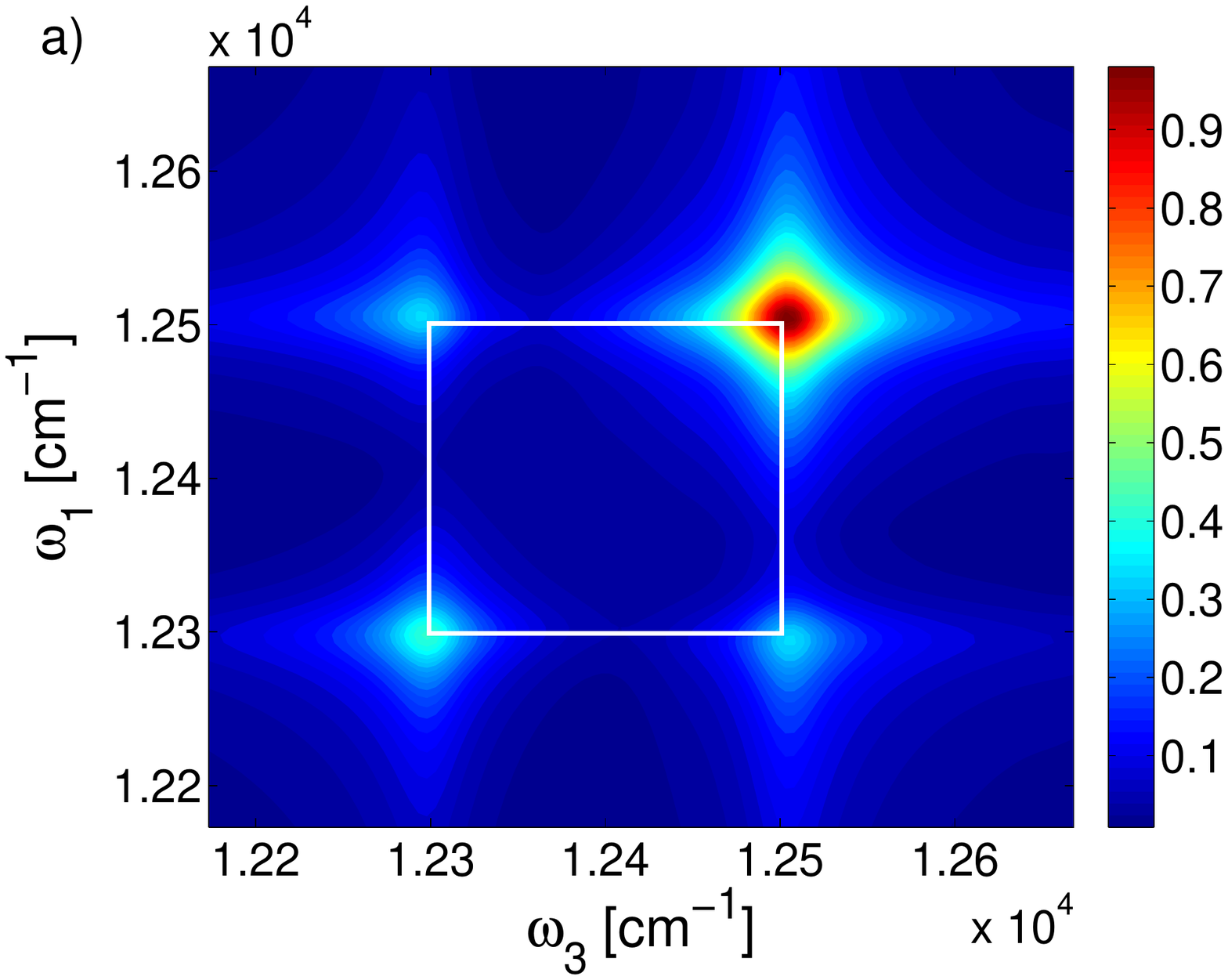}\\
\vspace*{-5.5cm}
\includegraphics[scale=0.37]{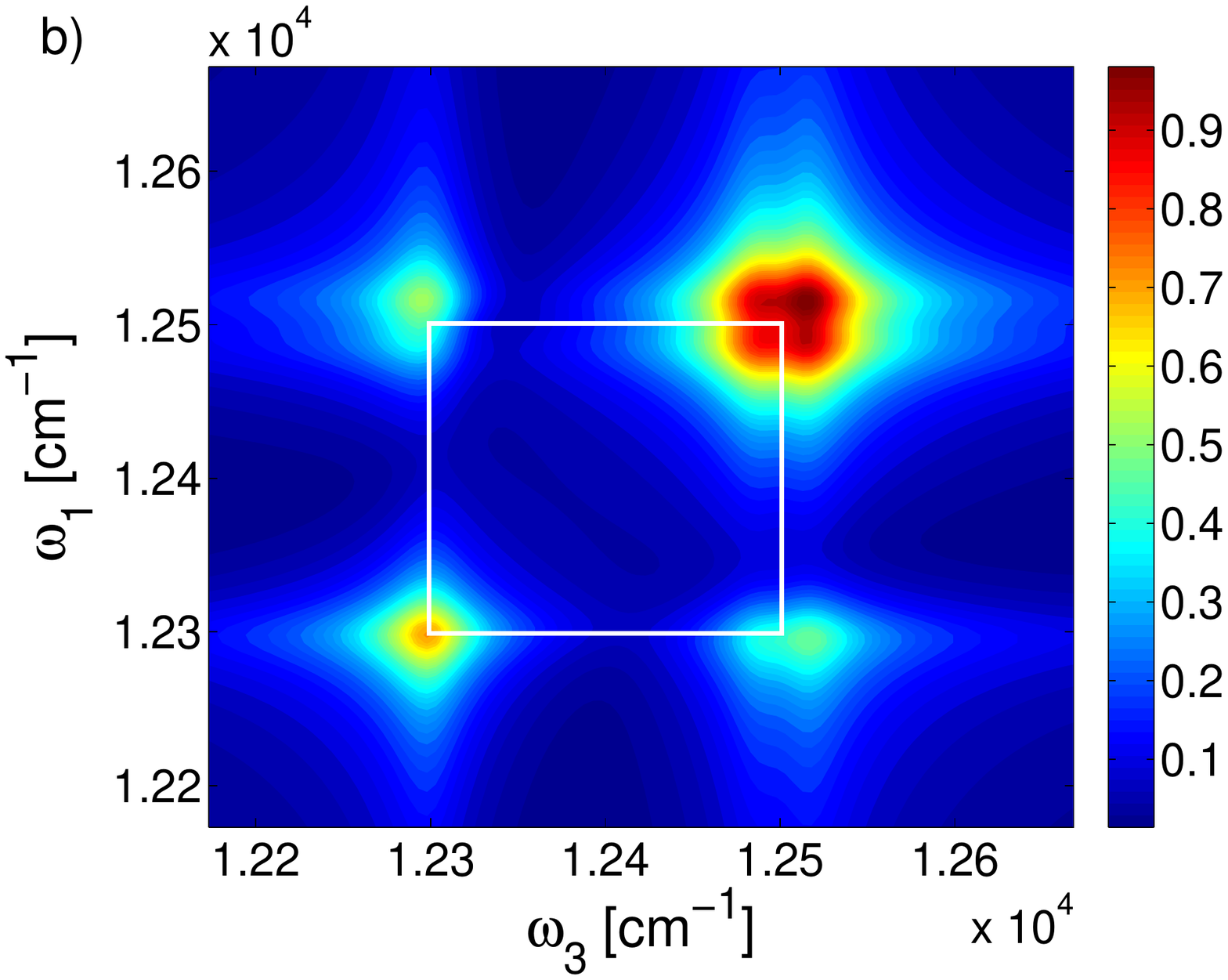}\\
\vspace*{-5.5cm}
\includegraphics[scale=0.37]{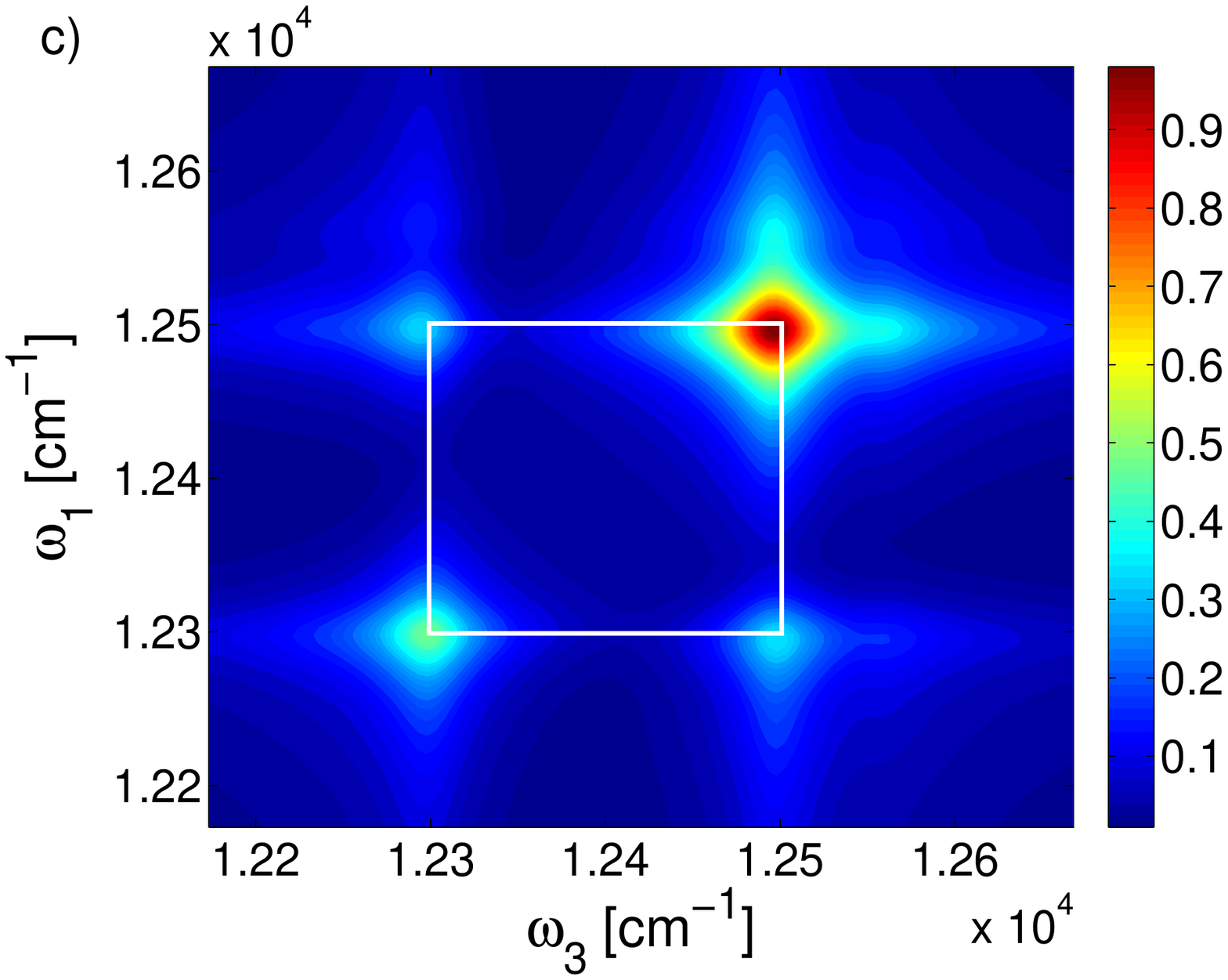}\\
\vspace*{-2.5cm}
\caption{\textbf{Resonant dipolar enhancement on the one-exciton sector.} Partial contribution ${\rm Abs}\{R_{3g}(\omega_1, t_2, \omega_3)\}$ of the total 2D electronic spectrum signal. The frequency of the mode in panels \textit{a}, \textit{b} and \textit{c} is $0.75\Delta_{\rm ex}$, $\Delta_{\rm ex}$ and $1.25\Delta_{\rm ex}$ respectively. The dipolar enhancement of the transitions to and from the hybridized states $\ket{H_{0,1}^{\pm}}$ is evident in the resonant case from the appearance of new resonances on each principal peak separated by a splitting that depends on the interaction strength with the mode. The environment temperature was set to $T=77K$ and the mode, initiated in a thermal state, suffers an inverse damping rate of $\gamma_{mode} = 0.005\Delta_{ex}$ into a reservoir at temperature $T$. The inverse dephasing rate for each site has been chosen as $\gamma_{\rm deph}= 0.025\Delta_{ex}$. The population time has been set to $t_2=32\Delta_{ex}^{-1}$ and the Huang-Rhys factor is $S_{\rm HR}=0.02$. The dipole moments of the two sites were chosen as $\vec{\mu}_a = (1,0.5,0)$ and $\vec{\mu}_b = (0,1,0)$ and the spectra include averaging over the orientation of the chromophores. All plots are normalized to the same maximal peak amplitude.}
\label{fig:diffpeaks}
\end{figure}
\noindent where $\mu_A$ and $\mu_B$ are the excitonic dipolar momenta and the mixing angle $\phi$ has some value close to $\pi/4$ near the resonance. This situation is illustrated in fig.~\ref{fig:dipolar_transitions}c, where now the state $\ket{H_0^{\pm}}$ can act as a {\em catalyzer} for vibrational transitions within the electronic ground state manifold. This particular redistribution allows the pathways described in fig.~\ref{fig:Feynman_diagrams}b to contribute to the total 2D spectra signal with strong amplitudes in the spirit described by Tiwari et al \cite{jonas}. Contrary to the non-resonant case, all these diagrams result in population time peak beating due to coherent superposition of states in the ground state electronic manifold. In particular, only the diagram corresponding to the cross-diagonal peak CP12 contains four electronic-vibrationally enhanced transitions. The diagrams corresponding to peaks DP2 and DP1 contain three enhanced transitions while the diagram corresponding to CP21 contains only two enhanced transitions. This intrinsic asymmetry between the different spectroscopic pathways has been postulated to be at the root of the differences found in actual experiments \cite{jonas}.
The redistribution of dipolar strength between the hybrid states $\ket{H^{\pm}_0}$ and $\ket{H^{\pm}_1}$ is indeed apparent in fig.~\ref{fig:diffpeaks}. In this figure we have plotted the $R_{3g}(\omega_1, t_2, \omega_3)$ diagram amplitude for different values of the harmonic mode frequency $\omega_m$ off- and on-resonance with the excitonic splitting $\Delta_{\rm ex}$. The traces of the hybridizations in the one-exciton manifold are clear in the on-resonance situation as now the four well resolved individual peaks corresponding to the excitonic frequencies split into a number of subpeaks which are the product of the self avoided level crossing due to electronic-vibrational coupling discussed above in fig. 1. The degree of splitting depends on the interaction strength with the vibrational mode.
%
\begin{figure}
\vspace*{-3.cm}
\includegraphics[scale=0.36]{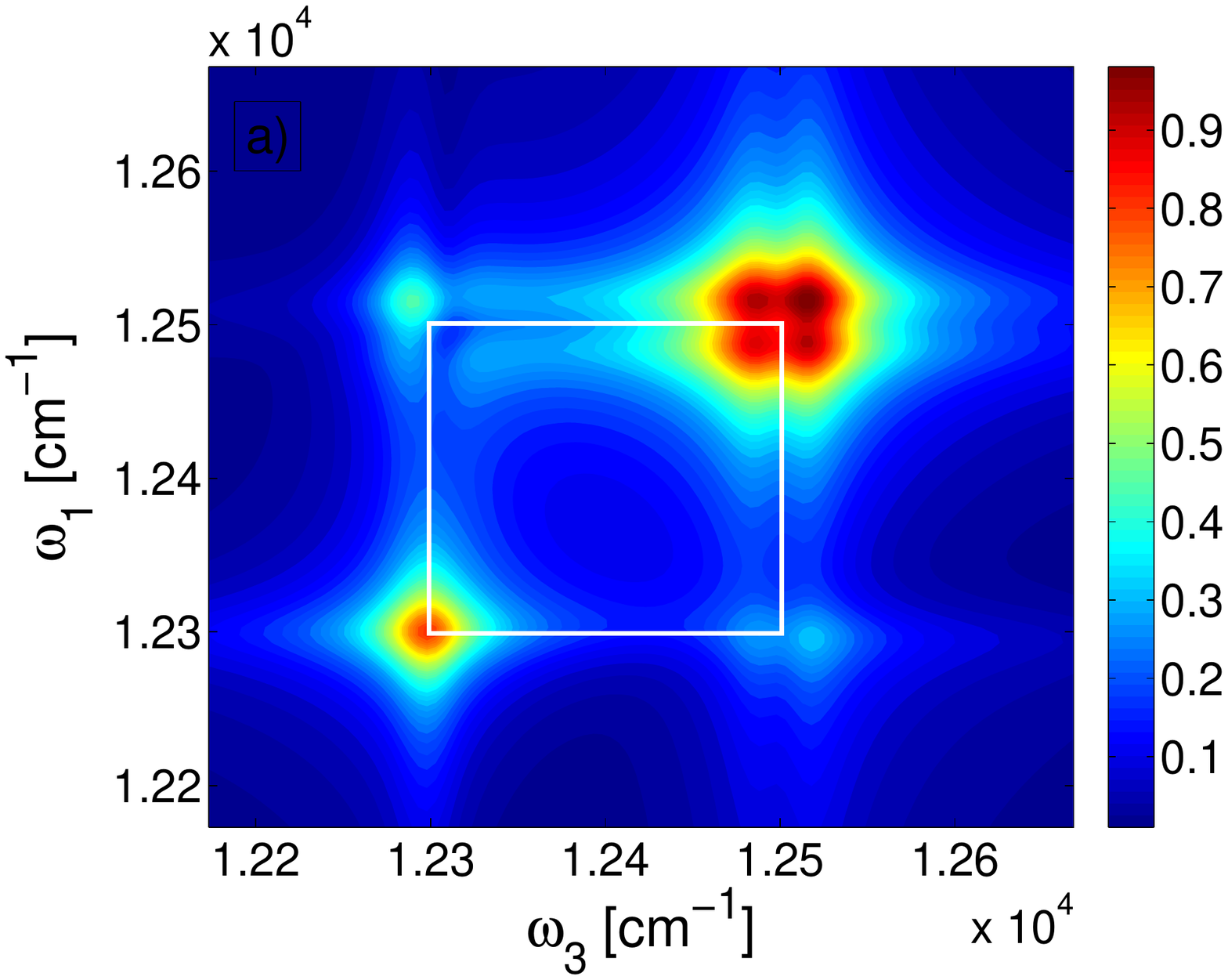}\\
\vspace*{-5.cm}
\includegraphics[scale=0.33]{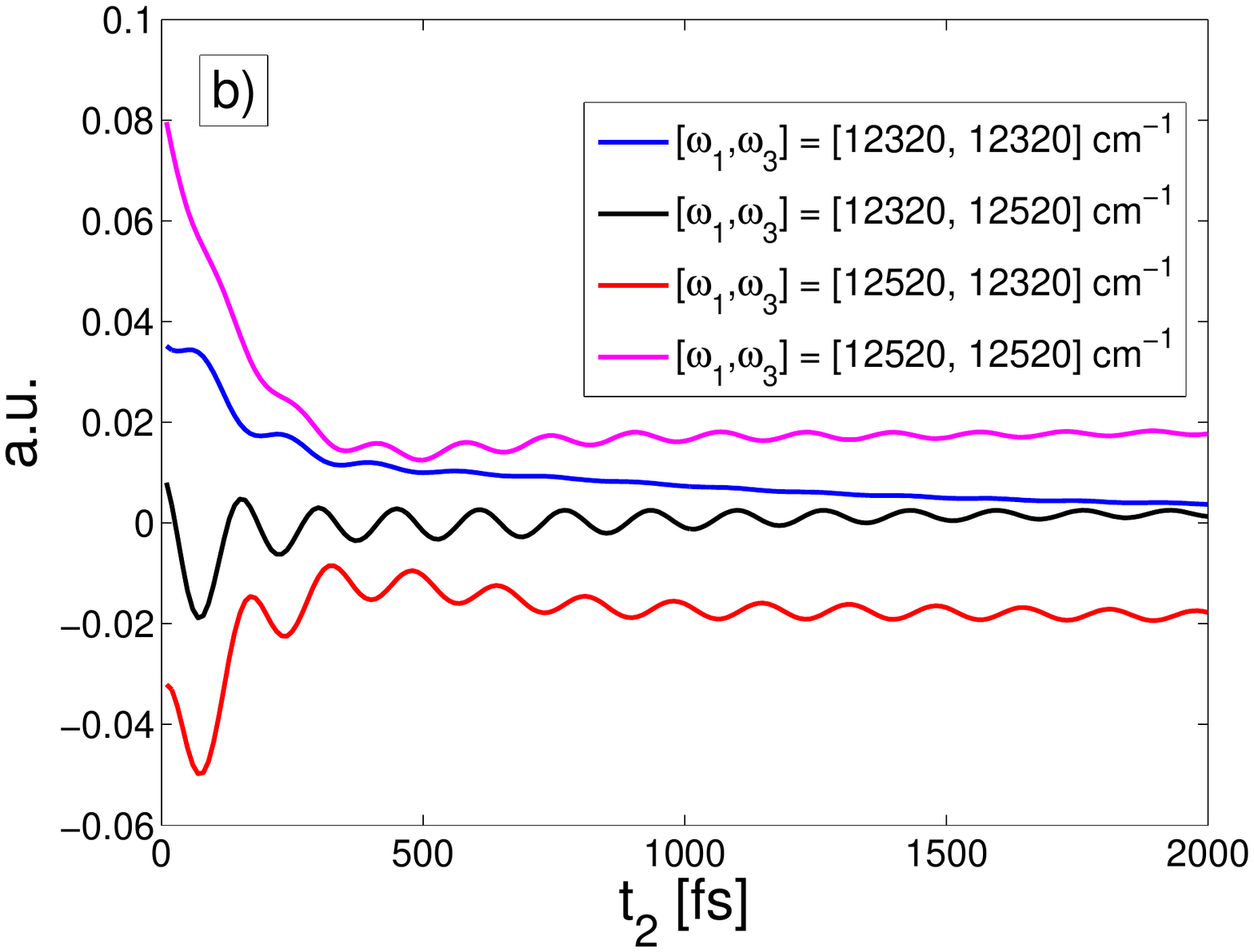}\\
\vspace*{-4.5cm}
\includegraphics[scale=0.33]{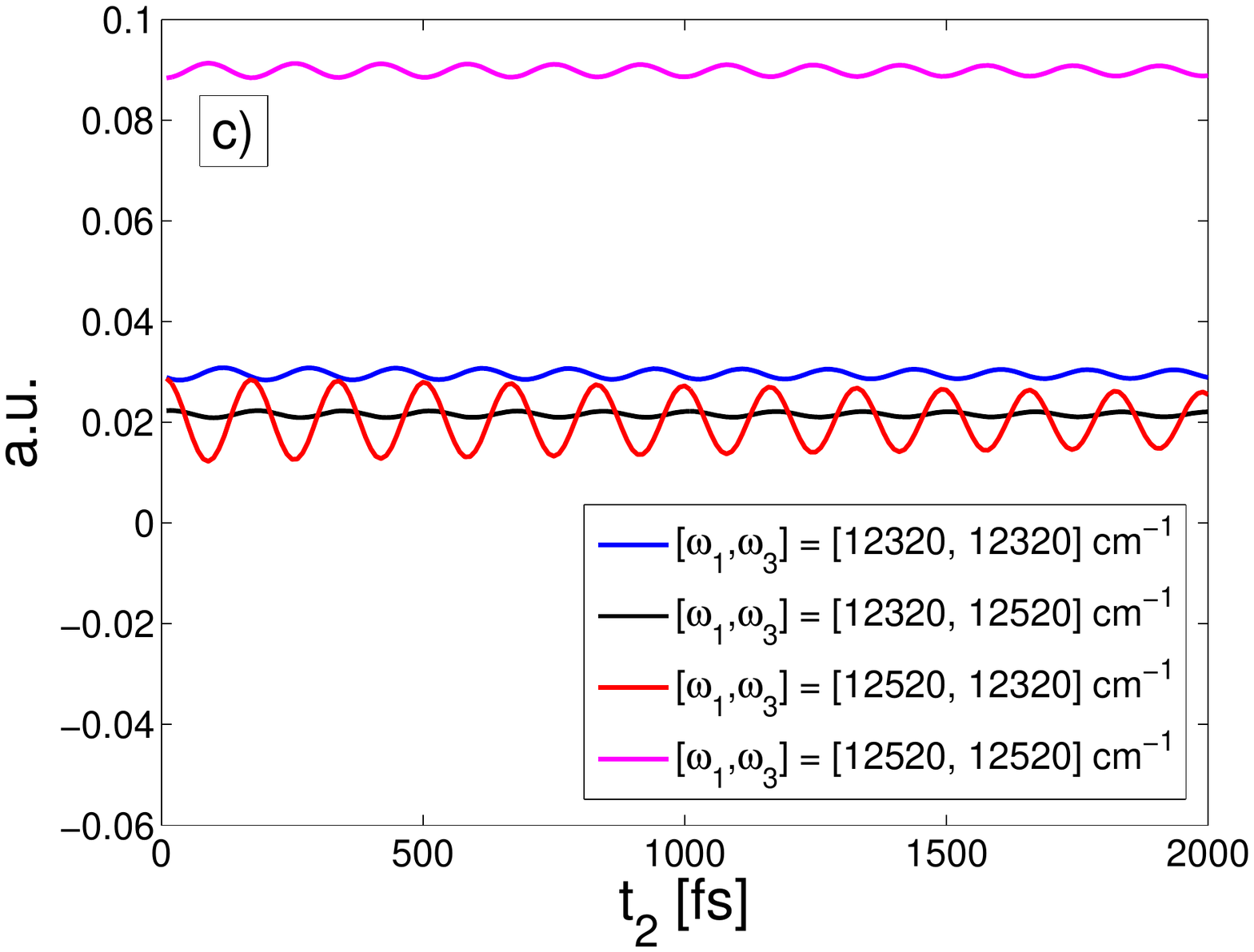}\\
\vspace*{-2.5cm}
\caption{\textbf{Electronic and vibrational population-time peak beating.}
\textit{a)} Absolute value of the electronic 2D spectra $-R^{*}_{1f} + R_{2g} + R_{3g}$ evaluated for a population time of $t_2 = 1.6\Delta_{ex}^{-1}$ and the vibrational modes on-resonance with the excitonic splitting $\Delta_{\rm ex}$. \textit{b)} Partial contribution ${\rm Re}(-R^{*}_{1f} + R_{2g})$ for the peaks (DP1 (blue), CP21 (black), CP12 (red), DP2 (magenta)) at $(\omega_1,\omega_3)=(12320,12320)$, $(12320,12520)$, $(12520,12320)$ and $(12520,12520)$. This contribution produces population-time peak beating given by coherences on the one-exciton sector (see diagrams in fig.~\ref{fig:Feynman_diagrams}a). \textit{c)} The partial contribution ${\rm Re}(R_{3g})$ for the same points. This contribution does not result in population-time beating on any peak in a purely electronic dimer. However, the mixing between electronic and vibrational degrees of freedom allows the transitions described in fig~\ref{fig:Feynman_diagrams}. The resulting beating in the population time is a signature of quantum coherence between vibrational states within the ground state electronic manifold. All other parameters are those of figure 4. Panels a), b) and c) include an average over the orientations of the dimer and panels b) and c) are also averaged over the static disorder with a probability density $p(\omega)\sim \exp^{-\omega^2/2\sigma^2}$ with $\sigma=0.17\Delta_{ex}$.}
\label{fig:beating_with_mode}
\end{figure}
\subsection{Ground state versus excited state contributions}
In what we have explained so far, the effect of the interaction between vibrational and electronic degrees of freedom is twofold. First it relates to 'lifetime borrowing' due to dynamics in the one-exciton sector and, secondly, it allows for the generation of coherent superpositions of vibrational states in the electronic ground state manifold through the excited state hybridization. Both phenomena contribute to the long lasting population time beating signals in 2D spectra. However, the intrinsic difference between the electronic sectors in which the relevant dynamics occurs in each case allows for a simple and transparent way to study their effects separately. According to fig.~\ref{fig:Feynman_diagrams}, the spectroscopic pathways whose dynamics is within the ground state sector during the population time are those described by $R_{3g}(\omega_1, t_2, \omega_3)$. On the other hand, the diagrams whose population time dynamics take place within the one-exciton sector are $R_{2g}(\omega_1, t_2, \omega_3)$ and
$R_{1f}^{*}(\omega_1, t_2, \omega_3)$. \\
\begin{figure}
\vspace*{-3.cm}
\includegraphics[scale=0.38]{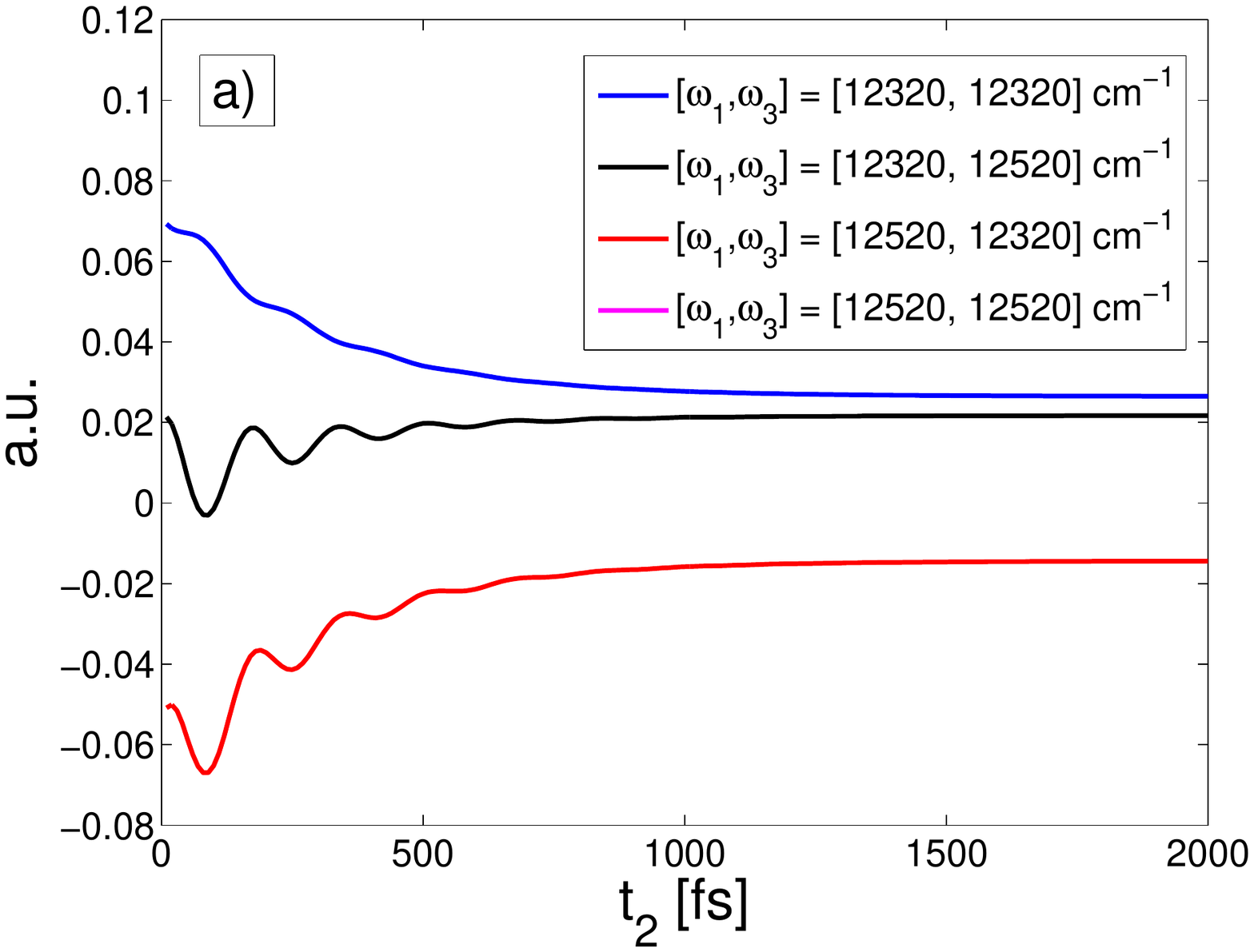}\\
\vspace*{-5.25cm}
\includegraphics[scale=0.38]{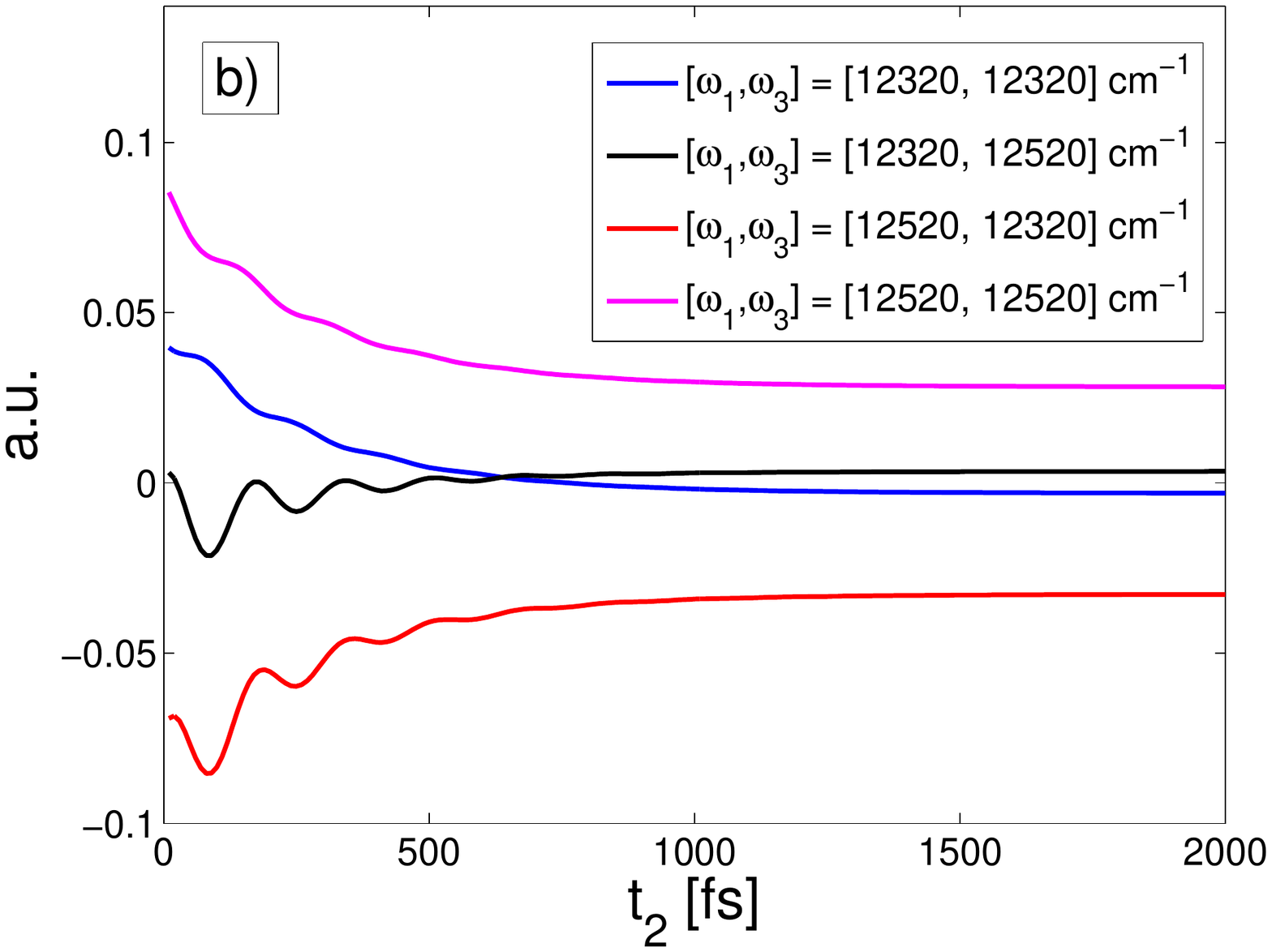}\\
\vspace*{-5.25cm}
\includegraphics[scale=0.38]{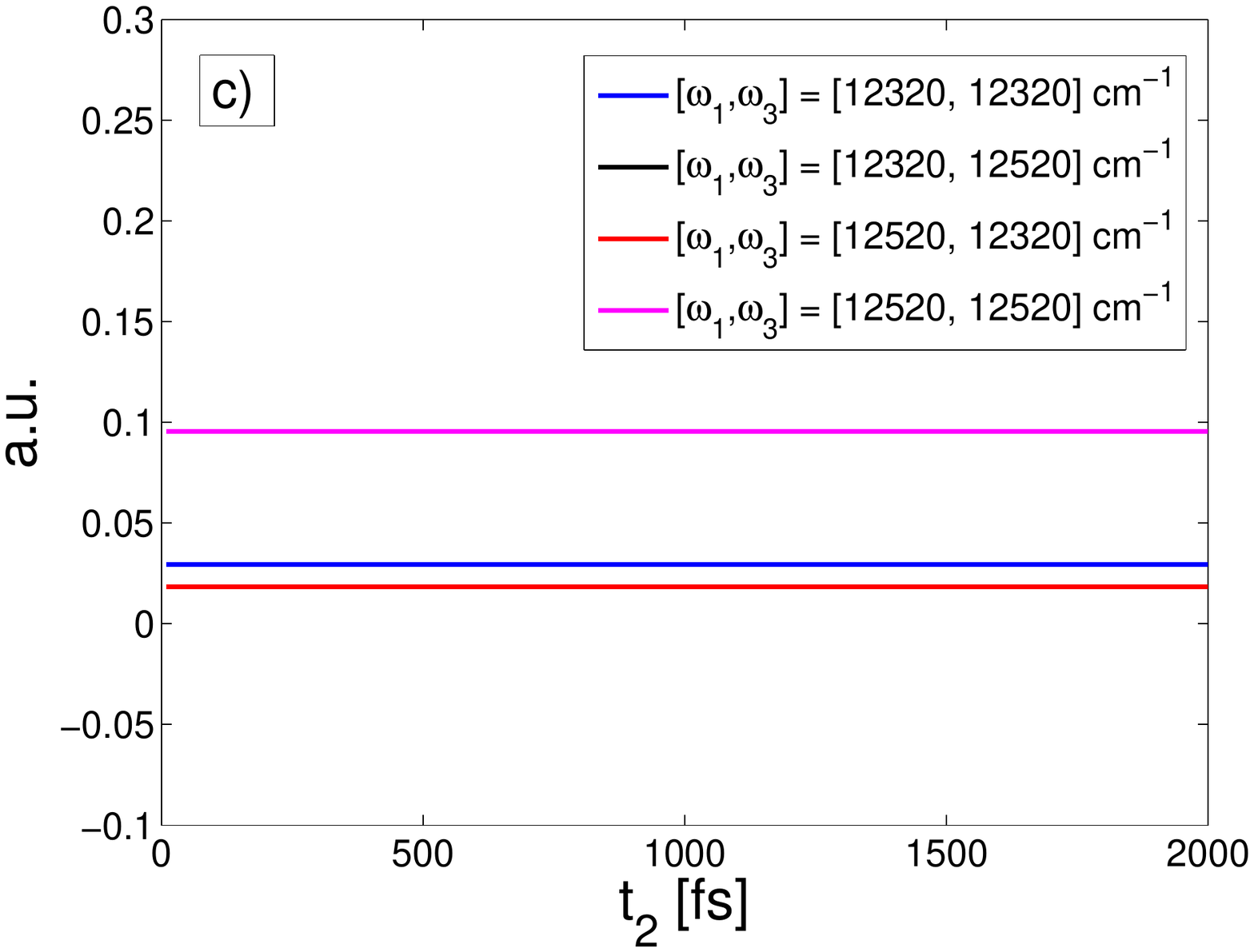}\\
\vspace*{-2.5cm}
\caption{\textbf{Purely electronic population-time peak beating} For a purely electronic dimer we study the population time beating for the same peaks of the 2D spectrum as in figure 5. Except for the Huang-Rhys factor which is now $S_{HR}=0$, all parameters and averaging is the same as in figures 5b and c. Notice that on a purely electronic model the diagram $Re(R_{3g}(\omega_1, t_2, \omega_3))$ does not exhibit any population-time beating neither on the cross- nor on the diagonal-peaks. On the other hand, the diagrams $Re(-R^{*}_{1f}(\omega_1, t_2, \omega_3) + R_{2g}(\omega_1, t_2, \omega_3))$ produce a beating only on the cross-peaks and not on the diagonal peaks. Indeed the modulation of the diagonal peaks above is not due to any excitonic superposition during the population time but due to the contamination with the nearby cross-diagonal peaks.}
\label{fig:beating_without_mode}
\end{figure}
In fig.~\ref{fig:beating_with_mode} we have plotted separately the contributions from these diagrams to the 2D spectrum in order to demonstrate that both are of comparable size. As we have seen, one of the first effects of a resonance between vibrational modes and excitonic energy differences is the appearance of a rich substructure on the principal excitonic peaks that are visible in the spectrum. In fig.~\ref{fig:beating_with_mode}a we show the full 2D spectrum, averaged over the random orientation of the complex, for an arbitrarily choice of the population time $t_2$ of the electronic-vibrational dimer where each cofactor is coupled to a vibrational mode as described by the Hamiltonian eq.(\ref{eq:Htot}). The first effect of a resonance between vibrational modes and excitonic energy differences is the appearance of a substructure on the principal excitonic peaks which splits into several peaks that are separated by the interaction strength between electronic and vibrational degrees of freedom. In general these peaks might have dipolar momenta with different signs arising from transitions to different hybridized states which leads to different relative phases between the peaks. The choice of the region to study is then arbitrary in a way, although we stress that the main conclusions, in particular those relating to the life-time of the population time beatings, are not affected by this choice.\\
In the panels fig.~\ref{fig:beating_with_mode}b and c we show the response of the system in the population time for the four peaks DP1, DP2, CP12 and CP21 depicted in fig.~\ref{fig:Feynman_diagrams}.b. Each of these time traces has been obtained from a simultaneous average over both the random orientation of the complex under investigation and over the unavoidable static disorder in the diagonal elements ${\cal E}_a$ and ${\cal E}_b$ of the Hamiltonian eq. \ref{eq:Hecd}. The computations have been carried out using an inverse dephasing rate for both chromophores equal to $\gamma_{deph}=0.025\Delta_{ex}$ and a harmonic mode that is itself damped to a Markovian environment with an inverse rate of $\gamma_{mode} = 0.005\Delta_{ex}$. The effect of the mechanisms described above are clearly observable in these graphs as in both cases the observed beatings exhibit a considerable extended life time when compared to the signals in the absence of the modes (see fig.~\ref{fig:beating_without_mode} for a comparison on the same electronic dimer decoupled from the vibrational mode).

\section{Conclusions} We have discussed the intricate interplay between electronic and vibrational degrees of freedom in the dynamics of molecular aggregates \cite{cp2013} and demonstrated how these results can play a crucial role in the dynamics of molecular aggregates and in particular in explaining the coherence properties of light harvesting complexes. Our results link recently proposed microscopic, non-equilibrium mechanisms in the electronic-vibrational coupling to support long-lived coherence in photosynthetic systems \cite{ourNP} with recent experimental observations of oscillatory behaviour using 2D photon echo techniques and clarify its relationship to recently proposed independent analysis \cite{jonas}. This analysis strongly suggests that the long-lived oscillations, and thus coherence, is vibrationally assisted but electronically facilitated, with the remarkable overall result of allowing for coherent oscillatory behaviour to be detectable on physiologically relevant time scales and at ambient temperature.

{\em Acknowledgements --} This work was supported by the Alexander von Humboldt-Foundation, the EU STREP project PAPETS and ERC Synergy grant BioQ. Aspects of this work have benefitted from discussions with F. Caycedo-Soler, A.~W. Chin and P. Fernandez-Acebal. We also acknowledge discussion with D.~M. Jonas and D. Miller at the KITP workshop on Quantum Control in February 2013 and with participants of the {\em Quantum transport in light-harvesting bio-nanostructures} meeting in Florence in March 2013 where this work was first presented.


\begin{thebibliography}{99}

\bibitem{greg1} G.~S. Engel, T. Calhoun, E. Read, T. Ahn, T. Man\v{c}al, Y. Cheng, R. Blankenship, and G.~R. Fleming, \newblock {\em Nature} {\bf 446}, 782 (2007).

\bibitem{greg2}
G. Panitchayangkoon, D. Hayes, K. Fransted, J. Caram, E. Harel, J. Wen, R. Blankenship, and G.~S. Engel,
\newblock {\em Proceedings of the National Academy of Sciences}{ \bf 107}, 12766 (2010).

\bibitem{collini} E. Collini, C. Wong, K. Wilk, P. Curmi, P. Brumer, and G.~D. Scholes, \newblock {\em Nature}{ \bf 463}(7281), 644 (2010).

\bibitem{vanhulst} R. Hildner, D. Brinks, J.~B. Nieder, R.~J. Cogdell, and N.~F. van Hulst, Science {\bf 340}, 1448 (2013).

\bibitem{newgreg} D. Hayes, G.~B. Griffin, and G.~S. Engel, Science {\bf 21}, 1431-1434 (2013).


\bibitem{felipe} F. Caycedo-Soler, A.~W. Chin, J. Almeida, S.~F. Huelga, and M.~B. Plenio, J. Chem. Phys. {\bf 136}, 155102 (2012).

\bibitem{mancal1} N. Christensson, H. Kauffmann, T. Pullerits, and T. Man\v{c}al, J. Phys. Chem. B, {\bf 116}, 7449 - 7454 (2012).

\bibitem{ourNP} A. W. Chin, J. Prior, R. Rosenbach, F. Caycedo-Soler, S.~F. Huelga, and M.~B. Plenio, Nature Physics {\bf 9}, 113 - 118 (2013).

\bibitem{jonas} V. Tiwari, W.~K. Peters, and D.~M. Jonas, Proc. Natl. Acad. Sci. U. S. A. {\bf 110}, 1203 - 1208 (2013).

\bibitem{mancalSR} A. Chenu, N. Christensson, H.~F. Kauffmann, and T. Man\v{c}al, Scientific Reports {\bf 3}, 2029 (2013).

\bibitem{cp2013}
S.~F. Huelga and M.~B. Plenio, {\em Vibrations, Quanta and Biology}, arXiv:1307.3530. In press Contemporary Physics (2013).

\bibitem{MohseniRL+08} M. Mohseni, P. Rebentrost, S. Lloyd, and A. Aspuru-Guzik,
J. Chem. Phys. {\bf 129}, 174106 (2008).

\bibitem{PlenioH08} M.~B. Plenio and S.F. Huelga,
New J. Phys. {\bf 10}, 113019 (2008).

\bibitem{CarusoCD+09} F. Caruso, A.~W. Chin, A. Datta, S.~F. Huelga, and M.~B. Plenio,
J. Chem. Phys. {\bf 131}, 105106 (2009).

\bibitem{ChinHP12} A.~W. Chin, S.~F. Huelga, and M.~B. Plenio.
Phil. Trans. Roy. Soc. A {\bf 370}, 3638 - 3657 (2012)

\bibitem{alexandra} A. Kolli,  E.~J. O'Reilly, G.~D. Scholes, and A. Olaya-Castro,
J. Chem. Phys. {\bf 137}, 174109 (2012).

\bibitem{delReyCH+13} M. del Rey, A.~W. Chin, S.~F. Huelga, and M.~B. Plenio,
J. Phys. Chem. Lett. {\bf 4}, 903 - 907 (2013).

\bibitem{scholes2011lessons} G.~D. Scholes, G.~R. Fleming, A. Olaya-Castro, and R. van Grondelle, \newblock {\em Nature Chemistry}{ \bf 3}(10), 763--774 (2011).

\bibitem{thomasreview} T. Renger and F. M\"{u}h, Phys.Chem.Chem.Phys. {\bf 15}, 24 (2013).

\bibitem{OlbrichSS+11a} C. Olbrich, J. Strumpfer, K. Schulten, and U. Kleinekath{\"o}fer,
J. Phys. Chem. Lett. {\bf 2}, 1771 - 1776 (2011).
%
\bibitem{OlbrichSS+11b} C. Olbrich, J. Strumpfer, K. Schulten, and U. Kleinekath{\"o}fer,
J. Phys. Chem. B {\bf 115}, 758 - 764 (2011).
%
\bibitem{ShimRV+12} S. Shim, P. Rebentrost, S. Valleau, and A. Aspuru-Guzik,
Biophys. J. {\bf 102}, 649 - 660 (2012).
%
\bibitem{RengerKS12} T. Renger, A. Klinger, F. Steinecker, M. Schmidt am Busch, J. Numata and F. M{\"u}h,
J. Phys. Chem. B {\bf 116}, 14565 - 14580 (2012).


\bibitem{marcus}
J. Adolphs, and T. Renger,
\newblock {\em Biophysical Journal}{ \bf 91}(8), 2778 (2006).

\bibitem{ratsep2007electron} M. R{\"a}tsep, and A. Freiberg,
\newblock {\em Journal of Luminescence}{ \bf 127}(1), 251--259 (2007).
%
\bibitem{fmo}
M.~T.~W. Milder, B. Br\"{u}ggemann, R. van Grondelle, and J.~L. Herek, Photosynth Res. {\bf 104}, 257–274 (2010).
%
%
\bibitem{echo}
D.~M. Jonas, Annu. Rev. Phys. Chem. {\bf 54}, 425–63 (2003).

\bibitem{echo1}
E.~L. Read, H. Lee, and G.~R. Fleming
Photosynth Res. {\bf 101}, 233–243 (2009).
%
\bibitem{echo2}
Readers with a quantum optical background may find useful the recent review by
A.~M. Branczyk, D.~B. Turner, and G.~D. Scholes, arXiv:1307.5855.
%
\bibitem{tonu}
P. Kjellberg, B. Br\"{u}ggemann, and T. Pullerits, Phys. Rev. B {\bf 74} 024303 (2006).
%
\end{thebibliography}
\end{document}